\documentclass{iopconfser}
\usepackage{graphicx} % Required for inserting images
\usepackage{amsmath,amssymb}
\usepackage{mathtools}
\usepackage{braket} %ブラケット
\usepackage{physics}
\usepackage{bm} %太字
\usepackage{mathrsfs}
\usepackage[version=3]{mhchem} %化学式
\usepackage{here}
\usepackage[bookmarks=true, bookmarksnumbered=true, colorlinks=true, linkcolor=blue, urlcolor=blue]{hyperref}
\usepackage{comment}
\usepackage{pdfpages}
\usepackage{multirow}
\usepackage[hang,small,bf]{caption}
\usepackage[subrefformat=parens]{subcaption}
\usepackage{float}
\usepackage{color}

\begin{document}

\title{Surface-localized topological superconductivity in nodal-loop materials: BdG analysis}

\author{Takeru Matsushima$^{1}$ and Hiroki Tsuchiura$^{1}$}

\affil{$^1$Department of Applied Physics, Graduate school of Engineering, Tohoku University, Sendai, Miyagi, Japan}
%\affil{$^2$Department, Institution, City, Country}

\email{matsushima.takeru.t3@dc.tohoku.ac.jp}

%%%%%%%%%%%%%%%%%%%%%%%%%%%%%%%%%%%%%%%%%%%%%%%%
\begin{abstract}
%%%%%%%%%%%%%%%%%%%%
We theoretically study surface superconductivity in a nodal-line
semimetal by combining a minimal tight-binding model with a
layer-resolved Bogoliubov--de~Gennes approach. In the normal state, the
model realizes a bulk nodal loop and an associated drumhead surface
band in a slab geometry with open boundaries in the $z$ direction: the
central layers reproduce the bulk-like density of states, whereas the
surface layer exhibits a sharp zero-energy peak originating from the
drumhead states. On top of this band structure we introduce chiral
$p$-wave and $d_{x^2-y^2}$-wave superconducting channels and determine
the layer-dependent gap amplitudes self-consistently. The chiral
$p$-wave order parameter is strongly enhanced at the outermost layers
and decays within only a few layers towards the interior, while the
$d$-wave order parameter is more than an order of magnitude smaller on
all layers. The quasiparticle dispersion and surface local density of
states in the chiral $p$-wave state show that the drumhead band is
efficiently gapped out and that the zero-energy peak in the normal
surface spectrum is split into two coherence peaks, directly reflecting
the induced superconducting gap. These results demonstrate that
superconductivity driven by drumhead surface states is naturally biased
toward a surface-localized chiral $p$-wave pairing symmetry and may
offer qualitative guidance for interpreting surface-sensitive
experiments on Pd-doped CaAgP.
%%%%%%%%%%%%%%%%%%%%
\end{abstract}
%%%%%%%%%%%%%%%%%%%%%%%%%%%%%%%%%%%%%%%%%%%%%%%%

%%%%%%%%%%%%%%%%%%%%%%%%%%%%%%%%%%%%%%%%%%%%%%%%
\section{Introduction}
%%%%%%%%%%%%%%%%%%%%%%%%%%%%%%%%%%%%%%%%%%%%%%%%
%Article text starts here. 
%Organise your text using section headings and include any equations, figures, tables, lists etc using your preferred LaTeX packages and commands.
%

Topological nodal-line semimetals have attracted considerable attention as a platform for unconventional superconductivity. In these systems, the bulk conduction and valence bands touch along closed loops in momentum space, and their surface projections support nearly flat ``drumhead'' states with a strongly enhanced density of states near the Fermi level. Such surface-localized bands are expected to be highly susceptible to interaction-driven instabilities, including magnetism and superconductivity, and therefore provide an ideal stage to explore exotic pairing states that are difficult to realize in conventional metals. Among candidate materials, the noncentrosymmetric pnictide CaAgP has been theoretically identified as a line-node Dirac semimetal (CaAgX, X = P, As), providing a concrete platform to explore drumhead-derived surface phenomena\cite{CaAgP_nodal}. Recent experiments on CaAgP and its Pd-doped derivatives, CaAg$_{1-x}$Pd$_x$P, have added further momentum to this direction: transport and point-contact measurements indicate the presence of high-mobility carriers consistent with a nodal-line semimetallic band structure and reveal superconducting transitions around 1--2 K that are strongly influenced by the surface states~\cite{CaAgP_transport,CaAgP_pointcontact}. Soft point-contact and tunneling spectroscopy have provided evidence that the superconductivity in Pd-doped CaAgP is unconventional, possibly surface-dominated and accompanied by broken time-reversal symmetry~\cite{CaAgP_TRSB1,CaAgP_TRSB2}. These findings naturally raise the question of which pairing symmetries are favored when superconductivity emerges from drumhead surface states in a nodal-line material. 
%\color{red}Related self-consistent theoretical work on surface superconductivity in Weyl semimetals has shown that, in a different topological semimetal setting, the bulk penetration of topological surface states can play an important role in determining surface superconductivity~\cite{Trama2025}.\color{black}

Stimulated by the theoretical proposal by Shapourian \textit{et al.} that nodal-loop materials may host topological crystalline superconductivity with chiral pairing symmetries~\cite{Shapourian2018}, we revisit the problem of superconductivity originating from drumhead surface states from a microscopic, self-consistent mean-field perspective. Our aim is more modest than a full topological classification: we focus on how the presence of drumhead states in a minimal nodal-line model biases the competition between different pairing symmetries, in particular between chiral $p$-wave and $d$-wave states that have both been discussed as candidates for unconventional superconductivity.
%=== added by HT on Mar. 18, 2026 =====
A related self-consistent study of surface superconductivity was recently reported for a time-reversal-symmetric Weyl semimetal with Fermi-arc surface states, where the momentum-dependent penetration of the surface states was shown to influence both the layer dependence of the order parameter and the momentum dependence of the superconducting gap at the surface\cite{Trama2025}.
%======================================

In this work, we study a minimal tight-binding model of a nodal-line semimetal in a slab geometry, described by the normal-state Hamiltonian $H_0$ in Eqs.~\eqref{eq:H0_slab}--\eqref{eq:Tz_def}. The model features a nodal line in the three-dimensional Brillouin zone and the associated drumhead surface states when open boundary conditions are imposed along the $z$ direction. On top of this normal-state band structure, we introduce superconducting pairing within a Bogoliubov--de Gennes (BdG) formalism, $H_{\rm BdG} = H_0 + H_\Delta$. We consider an odd-parity chiral $p$-wave channel and an even-parity $d_{x^2-y^2}$-wave channel, as a primary candidate, a chiral $p$-wave state with gap function $\Delta(\bm{k}_\parallel) = \Delta_0 (\sin k_x + i \sin k_y)\sigma_0$, which naturally couples to the in-plane momenta of the drumhead states. For comparison, we also examine a $d$-wave state with a sign-changing gap structure on the projected nodal-loop region. Rather than imposing a prescribed spatial profile, we determine the superconducting order parameters self-consistently in real space along the slab direction.

Our main goal is to clarify which pairing symmetry is favored when superconductivity develops predominantly from the drumhead surface states of a nodal-line semimetal, in a setting motivated by the experimental situation in CaAgP and Pd-doped CaAgP. By comparing the self-consistent solutions for chiral $p$-wave and $d$-wave order, we analyze (i) how the superconducting gap and quasiparticle spectrum evolve in the presence of surface-localized flat bands, (ii) how the local density of states on the surface reflects the opening of a gap in the drumhead band, and (iii) how the spatial profiles $\Delta_p(n)$ and $\Delta_d(n)$ depend on the layer index $n$. We show that superconductivity is strongly confined to the surface region, and that the chiral $p$-wave state is significantly more stable than the $d$-wave state in this setting, exhibiting an order parameter amplitude that is larger by orders of magnitude. These results highlight the crucial role of the drumhead states in selecting the pairing symmetry and provide a simple microscopic scenario in which surface-localized chiral $p$-wave superconductivity can naturally emerge in nodal-line materials such as CaAgP.

%%%%%%%%%%%%%%%%%%
\section{Model}
%%%%%%%%%%%%%%%%%%
We start from a minimal tight-binding model for a nodal-line semimetal
on a cubic lattice. Within the nearest-neighbor approximation, the
normal-state Hamiltonian in momentum space is written as
\begin{equation}
  \begin{split}
    H
    &=
    \sum_{\bm{k}}
    c^\dagger_{\bm{k}}
    \Bigl[
      \{-2(t_x \cos k_x + t_y \cos k_y + t_z \cos k_z) - M\}\,\sigma_1
      \\
      &\qquad\qquad\qquad\quad
      + 2 t_z \sin k_z\,\sigma_2
      +
      \{-\mu + t_q(2 - \cos k_x + \cos k_y)\}\,\sigma_0
    \Bigr]
    c_{\bm{k}} ,
  \end{split}
  \label{eq:H3D}
\end{equation}
where
$c^\dagger_{\bm{k}} = (c^\dagger_{\bm{k}\uparrow}, c^\dagger_{\bm{k}\downarrow})$
consists of two fermionic creation operators associated with a pseudospin degree of freedom, and $\sigma_{0,1,2}$ denote the
identity and Pauli matrices acting in pseudospin space. The physical electron spin is not included explicitly: in the $\ce{CaAgP}$-motivated context considered, since the spin–orbit coupling in CaAgP is negligible\cite{CaAgP_nodal}, the spin sector is treated as a trivial twofold degeneracy and is therefore suppressed throughout. The parameter $M$ is a mass term that controls the band inversion, $\mu$ is the chemical potential, and $t_x$, $t_y$, and $t_z$ are nearest-neighbor hopping
amplitudes along the $x$, $y$, and $z$ directions, respectively.
%The parameter $t_q$ introduces a weak momentum-dependent shift of the bands;
%in the calculations below we set $t_q=0$ for simplicity.
The parameter $t_q$ introduces a weak momentum-dependent shift of the bands.

To describe a slab geometry we impose open boundary conditions along the
$z$ direction and keep translational invariance in the in-plane
directions.  Introducing the in-plane momentum
$\bm{k}_{\parallel} = (k_x,k_y)$ and the layer index $n=1,\dots,N$, the
normal-state Hamiltonian can be written as
\begin{align}
  H_0
  &=
  \sum_{\bm{k}_{\parallel}}
  \Biggl[
    \sum_{n=1}^N
    c^\dagger_{\bm{k}_{\parallel},n}\,
    h_0(\bm{k}_{\parallel})\,
    c_{\bm{k}_{\parallel},n}
    +
    \sum_{n=1}^{N-1}
    \bigl(
      c^\dagger_{\bm{k}_{\parallel},n}\,T_z\,c_{\bm{k}_{\parallel},n+1}
      +
      c^\dagger_{\bm{k}_{\parallel},n+1}\,T_z^\dagger\,c_{\bm{k}_{\parallel},n}
    \bigr)
  \Biggr],
  \label{eq:H0_slab}
  \\
  h_0(\bm{k}_{\parallel})
  &=
  \{-2(t_x \cos k_x + t_y \cos k_y) - M\}\,\sigma_1
  +
  \{-\mu + t_q(2 - \cos k_x + \cos k_y)\}\,\sigma_0,
  \label{eq:h0_def}
  \\
  T_z
  &=
  \begin{pmatrix}
    0   & -2t_z \\
    0   & 0
  \end{pmatrix},
  \label{eq:Tz_def}
\end{align}
where
$c^\dagger_{\bm{k}_{\parallel},n}
 =(c^\dagger_{\bm{k}_{\parallel},n,\uparrow},
   c^\dagger_{\bm{k}_{\parallel},n,\downarrow})$
creates an electron in layer $n$.  The matrix $h_0(\bm{k}_{\parallel})$
describes the in-plane normal-state Hamiltonian within each layer, and
$T_z$ denotes the interlayer hopping matrix.  In the basis
\begin{equation}
  \bigl(
    c_{\bm{k}_{\parallel},1},
    c_{\bm{k}_{\parallel},2},
    \dots,
    c_{\bm{k}_{\parallel},N}
  \bigr)^{\mathrm{T}},
\end{equation}
the matrix part of $H_0$ is represented as
\begin{equation}
  \mathcal{H}_0(\bm{k}_{\parallel})
  =
  \begin{pmatrix}
    h_0         & T_z        & 0          & \cdots & 0 \\
    T_z^\dagger & h_0        & T_z        &        & \vdots \\
    0           & T_z^\dagger& h_0        & \ddots & 0 \\
    \vdots      &            & \ddots     & \ddots & T_z \\
    0           & \cdots     & 0          & T_z^\dagger & h_0
  \end{pmatrix}.
  \label{eq:H0_matrix}
\end{equation}

For the numerical calculations presented below we set
\(t_x = t_y = t_z \equiv t = 1\), \(t_q = 0\), and \(M = -3\).
Unless otherwise stated, the chemical potential is fixed at
\(\mu = 0\), so that the Fermi energy is taken as the zero of
energy. We consider a slab geometry with \(N = 20\) layers
along the \(z\) direction.
In the range $-4 < M < 4$ the two bands cross, and a nodal loop is
formed in the bulk band structure.  In particular, for $-4 < M < 0$ a
nodal loop appears around the $\Gamma$ point, and its radius shrinks as
$M$ approaches $-4$ from above.  When open boundary conditions are
imposed along $z$, the projection of the nodal loop onto the surface
Brillouin zone is accompanied by nearly flat drumhead surface states
inside the projected loop.  

We introduce an attractive interaction projected onto a single pairing channel, assumed to be local in the layer index and separable in momentum space:
\begin{equation}
  H_{\mathrm{int}} = -V\sum_{n=1}^N \sum_{\bm{k}_{\parallel}, \bm{k}'_{\parallel}} \phi(\bm{k}_{\parallel})\phi^*(\bm{k}'_{\parallel}) c^\dagger_{\bm{k}_{\parallel}, n}c^\dagger_{-\bm{k}_{\parallel},n}c_{-\bm{k}'_{\parallel},n}c_{\bm{k}'_{\parallel},n}, \quad (V>0).
\end{equation}
Under the mean-field decoupling, the pair potential is defined by
\begin{equation}
    \Delta_n(\bm{k}_{\parallel}) \equiv V \phi(\bm{k}_{\parallel}) \sum_{\bm{k}'_{\parallel}} \phi^*(\bm{k}'_{\parallel}) \ev{c_{-\bm{k}'_{\parallel}, n}c_{\bm{k}'_{\parallel}, n}} \equiv \Delta_{0, n}\phi(\bm{k_{\parallel}}),
\end{equation}
so that the coupling constant $V$ enters the self-consistent determination of the amplitude $\Delta_{0,n}$, while the momentum dependence is fixed by the chosen form factor $\phi(\bm{k}_{\parallel})$.

We now introduce superconductivity on top of the normal-state Hamiltonian
$H_0$.  The BdG Hamiltonian is written as
\begin{equation}
  H_{\mathrm{BdG}}
  =
  H_0 + H_{\Delta} ,
\end{equation}
where
\begin{align}
  H_{\Delta}
  &=
  \frac{1}{2}
  \sum_{\bm{k}_{\parallel}}
  \sum_{n=1}^N
  \sum_{\sigma, \sigma'}
  \Bigl[
    (\hat{\Delta}_{n}(\bm{k}_{\parallel}))_{\sigma,\sigma'}\,
    c^\dagger_{\bm{k}_{\parallel},n,\sigma}\,
    c^\dagger_{-\bm{k}_{\parallel},n,\sigma'}
    +
    \mathrm{h.c.}
  \Bigr].
\end{align}
Here $\phi(\bm{k_{\parallel}})$ is a form factor specifying the pairing symmetry (defined below). $\sigma$ denotes the same pseudospin index as in the normal-state Hamiltonian. We factorize the pair potential into a momentum form factor and an internal structure as,
\begin{equation}
  \hat{\Delta}_{n}(\bm{k}_{\parallel}) = \Delta_{0,n}\, \phi(\bm{k}_{\parallel})\,\Gamma.
\end{equation}
Here $\Gamma$ specifies the pairing channel in the pseudospin space. In this work we use $\Gamma = i\sigma_2$ for the even-parity $d_{x^2-y^2}$-wave form factor with an antisymmetric internal structure, while $\Gamma$ is $\sigma_0$ for the odd-parity chiral $p$-wave channel. The gap matrix satisfies $\hat{\Delta}_n(\bm{k}_{\parallel})=-\hat{\Delta}_n^{\mathrm{T}}(-\bm{k}_{\parallel})$; hence an even-parity form factor requires an antisymmetric $\Gamma \,\,(\Gamma=i\sigma_2)$, while an odd-parity form factor allows a symmetric $\Gamma\,\,(\mathrm{e.g.}, \Gamma=\sigma_0)$.

To formulate the BdG problem we introduce the Nambu spinor
\begin{equation}
  \Psi_{\bm{k}_{\parallel},n}
  =
  \begin{pmatrix}
    c_{\bm{k}_{\parallel},n,\uparrow} \\
    c_{\bm{k}_{\parallel},n,\downarrow} \\
    c^\dagger_{-\bm{k}_{\parallel},n,\uparrow} \\
    c^\dagger_{-\bm{k}_{\parallel},n,\downarrow}
  \end{pmatrix},
  \qquad
  \Psi_{\bm{k}_{\parallel}}
  =
  \bigl(
    \Psi_{\bm{k}_{\parallel},1},
    \Psi_{\bm{k}_{\parallel},2},
    \dots,
    \Psi_{\bm{k}_{\parallel},N}
  \bigr)^{\mathrm{T}} ,
\end{equation}
so that the BdG Hamiltonian can be written as
\begin{align}
  H_{\mathrm{BdG}}
  &=
  \frac{1}{2}
  \sum_{\bm{k}_{\parallel}}
  \Psi^\dagger_{\bm{k}_{\parallel}}\,
  \mathcal{H}_{\mathrm{BdG}}(\bm{k}_{\parallel})\,
  \Psi_{\bm{k}_{\parallel}},
  \\
  \mathcal{H}_{\mathrm{BdG}}(\bm{k}_{\parallel})
  &=
  \begin{pmatrix}
  \mathcal{H}_0(\bm{k}_{\parallel})
    &
    \hat{\Delta}(\bm{k}_{\parallel})
    \\
    \hat{\Delta}^\dagger(\bm{k}_{\parallel})
    &
    -\mathcal{H}^{\mathrm{T}}_0(-\bm{k}_{\parallel})
  \end{pmatrix},
  \label{eq:Hbdg_matrix}
\end{align}
here,
\begin{equation}
  \hat{\Delta}(\bm{k}_{\parallel})=
  \begin{pmatrix}
    \hat{\Delta}_1(\bm{k}_{\parallel})      & 0        & 0          & \cdots & 0 \\
    0 & \hat{\Delta}_2(\bm{k}_{\parallel})        & 0        &        & \vdots \\
    0           & 0& \hat{\Delta}_3(\bm{k}_{\parallel})        & \ddots & 0 \\
    \vdots      &            & \ddots     & \ddots & 0 \\
    0           & \cdots     & 0          & 0 & \hat{\Delta}_N(\bm{k}_{\parallel})
  \end{pmatrix}.
\end{equation}
Thus $\mathcal{H}_{\mathrm{BdG}}(\bm{k}_{\parallel})$ is represented as a
block-tridiagonal matrix in the layer space.

$\Delta_{0,n}$ is a complex gap amplitude on layer $n$, and
$\phi(\bm{k}_{\parallel})$ is a form factor that encodes the pairing
symmetry.  In the following we focus on a single pairing channel at a
time and choose
\begin{equation}
  \phi(\bm{k}_{\parallel})
  =
  \sin k_x + i \sin k_y
  \quad \text{(chiral $p$-wave)},
\end{equation}
or
\begin{equation}
  \phi(\bm{k}_{\parallel})
  =
  \cos k_x - \cos k_y
  \quad \text{($d_{x^2-y^2}$-wave)},
\end{equation}
depending on the case under consideration.

Diagonalizing $\mathcal{H}_{\mathrm{BdG}}(\bm{k}_{\parallel})$ yields
quasiparticle energies $E_\nu(\bm{k}_{\parallel}) \ge 0$ and eigenvectors
\begin{equation}
  \Phi_{\nu}(\bm{k}_{\parallel})
  =
  \begin{pmatrix}
    u_{1\nu\uparrow}(\bm{k}_{\parallel}) \\
    u_{1\nu\downarrow}(\bm{k}_{\parallel}) \\
    \vdots \\
    u_{N\nu\downarrow}(\bm{k}_{\parallel}) \\
    v_{1\nu\uparrow}(\bm{k}_{\parallel}) \\
    \vdots \\
    v_{N\nu\downarrow}(\bm{k}_{\parallel})
  \end{pmatrix},
\end{equation}
where $u_{n\nu\sigma}(\bm{k}_{\parallel})$ and
$v_{n\nu\sigma}(\bm{k}_{\parallel})$ are, respectively, the electron and
hole components on layer $n$ with internal index $\sigma$ in the
quasiparticle state labeled by $\nu$.  Restricting the sum over $\nu$ to
positive-energy branches, the self-consistent gap equation on layer $n$
can be expressed as
\begin{equation}
  \Delta_{0,n}
  =
  V
  \sum_{\bm{k}_{\parallel}}
  \phi^*(\bm{k}_{\parallel})
  \sum_{\nu>0}\sum_{\sigma,\sigma'}
  \Gamma_{\sigma\sigma'}
  u_{n\nu\sigma}(\bm{k}_{\parallel})\,
  v_{n\nu\sigma'}^*(\bm{k}_{\parallel})\,
  \tanh\!\left(
    \frac{E_\nu(\bm{k}_{\parallel})}{2T}
  \right).
  \label{eq:gap_eq}
\end{equation}
The contraction with $\Gamma$ selects the pairing channel (e.g., antisymmetric vs. symmetric).
In this formulation the layer dependence enters through the amplitudes $\Delta_{0,n}$ and the BdG eigenvectors, while the choice of $\phi(\bm{k}_{\parallel})$ selects the chiral $p$-wave or $d_{x^2-y^2}$-wave pairing channel.  For the calculations below we set $T=0$ and $V=0.5$.  All numerical computations are performed using a Python-based code, and the self-consistency iterations for Eq.~\eqref{eq:gap_eq} are terminated when the relative change of $|\Delta_{0,n}|/t$ falls below $10^{-5}$. The Brillouin zone is discretized using a \(100 \times 100\) mesh for \((k_x, k_y)\). The same $\bm{k}_\parallel$ mesh is used for all local density of states (LDOS) calculations shown below, including the open boundary conditions (OBC) slab calculation in Fig. \ref{fig:normal}\subref{fig:LDOS_normal}, as well as the OBC slab data in Fig. \ref{fig:SC}\subref{fig:LDOS_SC_3} and \ref{fig:SC}\subref{fig:LDOS_SC_0.25}. In Fig. \ref{fig:normal}\subref{fig:LDOS_normal}, the dataset labeled “unitary” corresponds to the fully periodic three-dimensional $\bm{k}$-space calculation, and only this dataset is obtained using a $200 \times 200$ mesh.

To analyze the spectral properties we compute the layer- and
momentum-resolved retarded Green's function in the superconducting state.
We compute the retarded BdG Green's function as
\begin{equation}
  G^R(\bm{k}_{\parallel},E)
  =
  \bigl[
    (E + i\eta)\,\hat{1}
    - 
    \mathcal{H}_{\mathrm{BdG}}(\bm{k}_{\parallel})
  \bigr]^{-1} \label{eq:Green function}.
\end{equation}
In the following we use a small positive broadening $\eta$ in Eq.~\eqref{eq:Green function} and set $\eta/t = 0.05$. We define the internal-space Green's function on layer $n$ as
\begin{equation}
  G_n^R(\bm{k}_{\parallel},E)
  =
  \begin{pmatrix}
    G_{n\uparrow,n\uparrow}^R(\bm{k}_{\parallel},E)
      & G_{n\uparrow,n\downarrow}^R(\bm{k}_{\parallel},E) \\
    G_{n\downarrow,n\uparrow}^R(\bm{k}_{\parallel},E)
      & G_{n\downarrow,n\downarrow}^R(\bm{k}_{\parallel},E)
  \end{pmatrix},
\end{equation}
where
$G^R_{n\sigma,n\sigma'} = \bigl[G^R_{ee}\bigr]_{n\sigma,n\sigma'}$
and $G^R_{ee}$ is the electron–electron block of the full BdG Green's
function.  The layer-, momentum-, and energy-resolved LDOS is then given by
\begin{equation}
  D_n(\bm{k}_{\parallel},E)
  =
  -\frac{1}{\pi}\,
  \Im\Bigl[
    G_{n\uparrow,n\uparrow}^R(\bm{k}_{\parallel},E)
    +
    G_{n\downarrow,n\downarrow}^R(\bm{k}_{\parallel},E)
  \Bigr].
  \label{eq:ar-LDOS}
\end{equation}
The layer-resolved LDOS is obtained by summing over in-plane momenta,
\begin{equation}
  D_n(E)
  =
  \frac{1}{N_{\parallel}}
  \sum_{\bm{k}_{\parallel}}
  D_n(\bm{k}_{\parallel},E),
\end{equation}
where $N_{\parallel}$ is the number of $\bm{k}_{\parallel}$ points in
the Brillouin zone.

For comparison, the normal-state LDOS is evaluated by using the same
expression, but with the BdG Hamiltonian replaced by the normal-state
Hamiltonian,
\begin{equation}
  \mathcal{H}_{\mathrm{BdG}}(\bm{k}_{\parallel})
  \;\longrightarrow\;
  \mathcal{H}_0(\bm{k}_{\parallel})
  \qquad (\Delta = 0),
\end{equation}
so that $G_n^R(\bm{k}_{\parallel},E)$ is computed from the normal-state
Green's function.  In the band-structure plots of
Figs.~\ref{fig:normal}\subref{fig:band_PBC} and \ref{fig:normal}\subref{fig:band_OBC}, the weight on layer $n$ in a quasiparticle band $\nu$ is defined as
\begin{equation}
  w_{n\nu}(\bm{k}_{\parallel})
  =
  \sum_{\sigma}
  \Bigl(
    |u_{n\nu\sigma}(\bm{k}_{\parallel})|^2
    +
    |v_{n\nu\sigma}(\bm{k}_{\parallel})|^2
  \Bigr),
\end{equation}
and the color map is drawn according to the normalized quantity
\begin{equation}
  \frac{w_{n\nu}(k_x)}{\max_{\nu} w_{n\nu}(k_x)}
  \in [0,1].
\end{equation}

%%%%%%%%%%%%%%%%%%%%%
\section{Results}
%%%%%%%%%%%%%%%%%%%%%
\subsection{Normal-state band structure and drumhead surface states}

First, let us confirm that, in the normal state, the nodal-line model in a slab geometry indeed hosts a bulk nodal loop and the associated drumhead surface states.
Throughout this section we set $t_x=t_y=t_z\equiv t=1$ and $M=-3$, and
consider a slab geometry with $N=20$ layers, consistent with the model
introduced in Sec.~2. The Fermi energy is taken as the zero of energy.

\begin{figure}[t]
\centering
\begin{minipage}[t]{0.3\columnwidth}
    \centering
    \includegraphics[width=0.9\columnwidth]{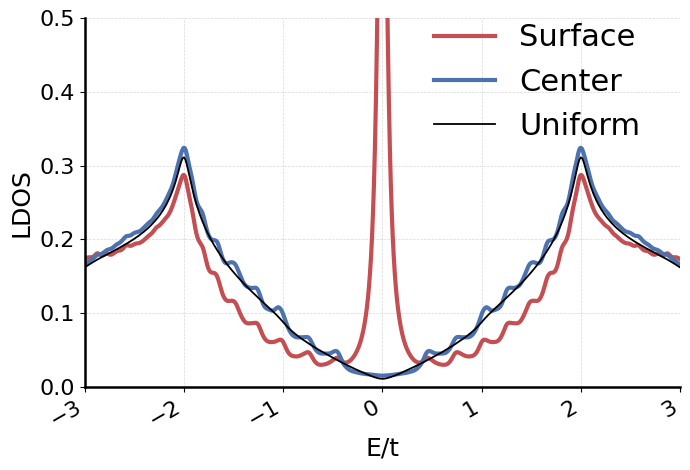}
    \subcaption{}
    \label{fig:LDOS_normal}
\end{minipage}
\begin{minipage}[t]{0.3\columnwidth}
    \centering
    \includegraphics[width=0.9\columnwidth]{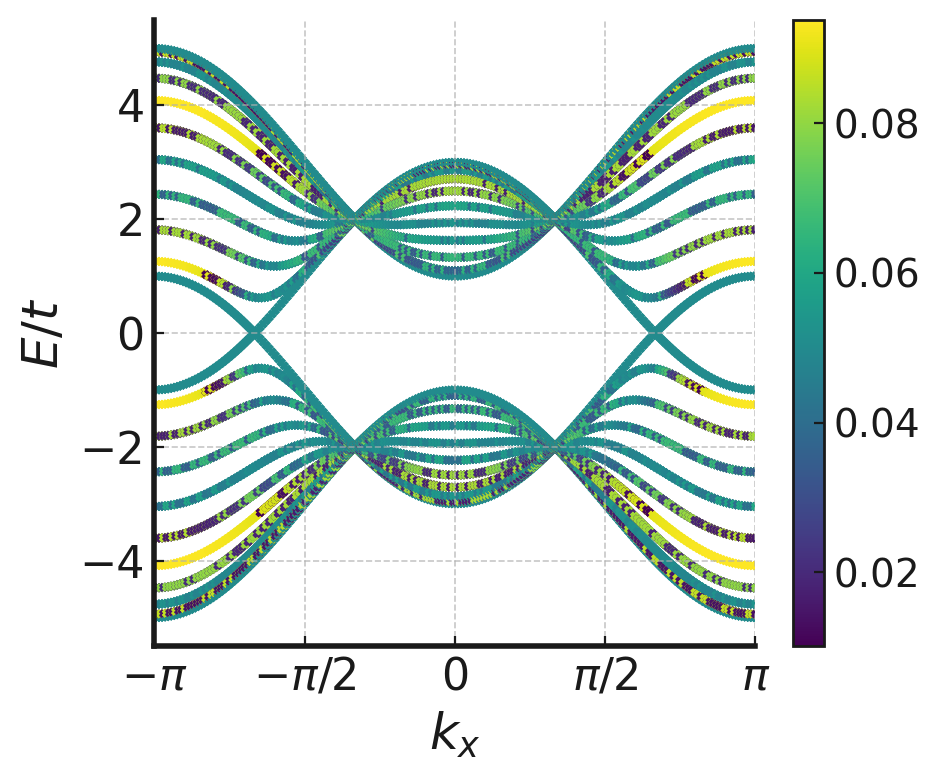}
    \subcaption{}
    \label{fig:band_PBC}
\end{minipage}
\begin{minipage}[t]{0.3\columnwidth}
    \centering
    \includegraphics[width=0.9\columnwidth]{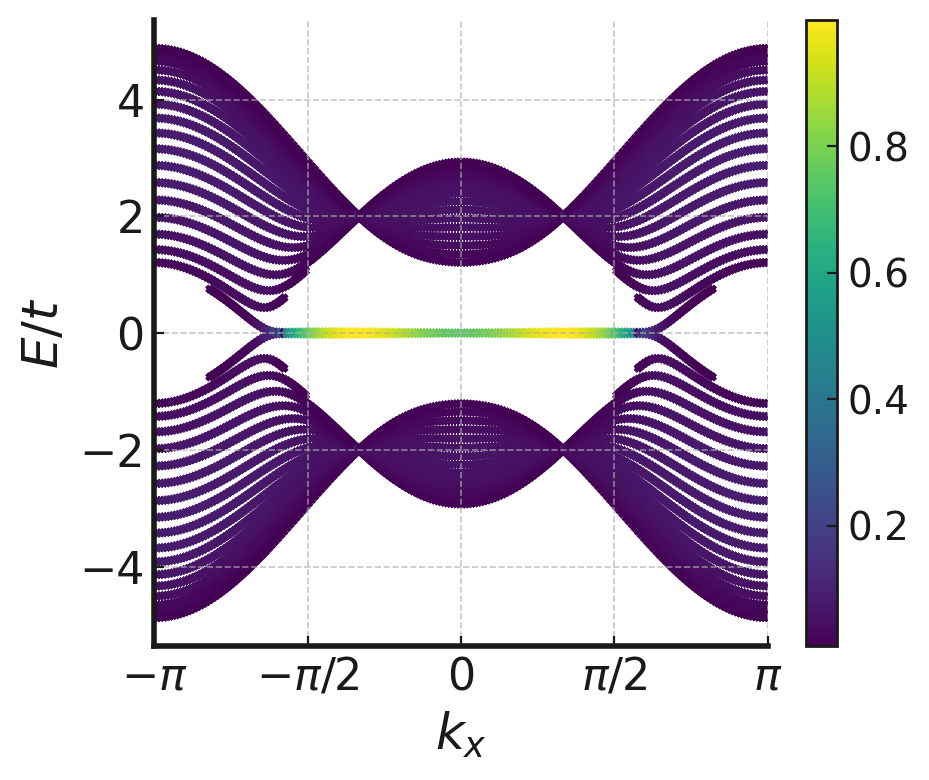}
    \subcaption{}
    \label{fig:band_OBC}
\end{minipage}
\caption{
Normal-state properties of the nodal-line model.
(a) LDOS in the fully periodic three-dimensional $k$-space model (thin black line),
together with the LDOS at the central layer (blue line) and at the surface layer (red line)
of a 20-layer slab with open boundary conditions in the $z$ direction.
The close agreement between the black and blue curves confirms that the central layer is bulk-like,
whereas the red curve exhibits a pronounced and sharp peak near $E=0$ originating from the surface states.
(b) Band dispersion $E(k_x)$ along $k_y=0$ for the fully periodic system, showing the crossings near
$k_x \simeq \pm 2\pi/3$ associated with the bulk nodal loop.
(c) Band dispersion for the slab with open boundary conditions in the $z$ direction along $k_y=0$,
illustrating the almost flat drumhead surface band at $E=0$ in the interval
$-2\pi/3 \lesssim k_x \lesssim 2\pi/3$.
} \label{fig:normal}
\end{figure}

To verify that the slab faithfully captures the bulk electronic structure in its interior and simultaneously hosts surface-localized states, we compare the LDOS of the fully periodic three-dimensional model with the layer-resolved LDOS of an $N=20$ slab with OBC in the $z$ direction.
As shown in Fig.\ref{fig:normal}\subref{fig:LDOS_normal}, the LDOS at the central layer ($n=10$, blue line) closely follows that of the fully periodic $k$-space model (thin black line), indicating that the central region is bulk-like to a good approximation.
In contrast, the LDOS at the surface layer ($n=1$, red line) exhibits a pronounced and sharp peak near $E=0$, signaling the presence of surface-localized states.
%%%%%%%%%%%%%%%%%%%%%%%%%%%%%%%%%%%%%%5
%%%%%%%%%%%%%%%%%%%%%%%%%%%%%%%%%%%%%%5
The normal-state band dispersion for the fully periodic system is presented in Fig.\ref{fig:normal}\subref{fig:band_PBC}, where the energy is plotted as a function of $k_x$ along the cut $k_y=0$.
At the Fermi level $E=0$, the bands cross near $k_x \simeq \pm 2\pi/3$, showing that the nodal loop of the three-dimensional spectrum intersects this momentum cut at these points.
This confirms that the chosen parameter set realizes a nodal-line semimetal.

We now turn to the slab geometry with OBC in the $z$ direction.
The corresponding band dispersion is shown in Fig.\ref{fig:normal}\subref{fig:band_OBC}, again along the cut $k_y=0$.
In addition to the bulk-like bands, an almost dispersionless band appears at $E=0$ in the interval $-2\pi/3 \lesssim k_x \lesssim 2\pi/3$.
This nearly flat band accounts for the zero-energy peak in the surface LDOS in Fig.\ref{fig:normal}\subref{fig:LDOS_normal} and is naturally identified as the drumhead surface band arising from the bulk nodal loop.
In the following, we study how superconductivity develops on top of this normal-state band structure and, in particular, how the drumhead surface states influence the stability and spatial profile of different pairing symmetries.

%%%%%%%%%%%%%%%%%%%%%%%%%%%%%%%%%%%%%%%%%%%%%%%%%%%%%%%%%%%%%%%%%%%%%%%%
\subsection{Superconducting states: chiral $p$- and $d$-wave pairing}
%%%%%%%%%%%%%%%%%%%%%%%%%%%%%%%%%%%%%%%%%%%%%%%%%%%%%%%%%%%%%%%%%%%%%%%%
We now investigate superconducting states that develop on top of the
nodal-line band structure within the Bogoliubov--de~Gennes framework
introduced in Sec.~2. We focus on a single pairing channel with a factorized gap function
$\Delta_n(\bm{k}_\parallel)=\Delta_{0,n}\,\phi(\bm{k}_\parallel)$ and
consider two representative form factors:
$\phi(\bm{k}_\parallel)=\sin k_x + i\sin k_y$ for a chiral $p$-wave
state and $\phi(\bm{k}_\parallel)=\cos k_x - \cos k_y$ for a
$d_{x^2-y^2}$-wave state. For a given pairing channel, the
layer-dependent amplitudes $\Delta_{0,n}$ are obtained by solving the
self-consistent gap equation~\eqref{eq:gap_eq} at $T=0$ for a fixed
attractive interaction $V$.

\begin{figure}[t]
    \centering
    \includegraphics[width=0.9\columnwidth]{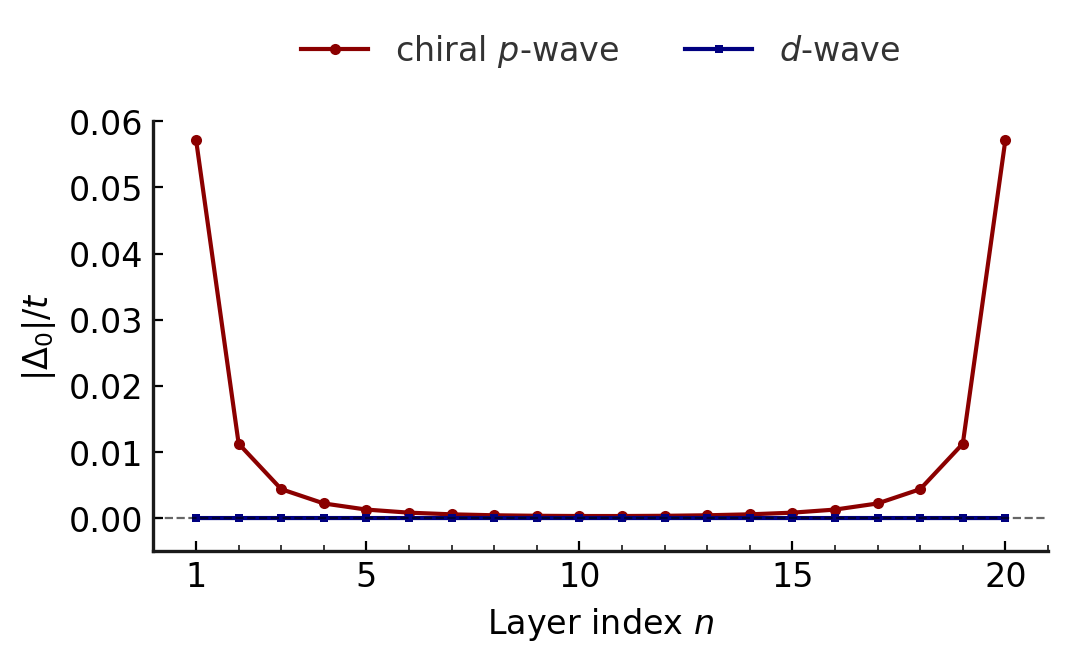}
%    \subcaption{}
%    \label{fig:delta}
%\centering
%\begin{minipage}[t]{0.45\columnwidth}
%    \centering
%    \includegraphics[width=0.9\columnwidth]{delta.png}
%    \subcaption{}
%    \label{fig:delta}
%\end{minipage}
%\begin{minipage}[t]{0.45\columnwidth}
%    \centering
%    \includegraphics[width=0.9\columnwidth]{delta_log.png}
%    \subcaption{}
%    \label{fig:delta_log}
%\end{minipage}
\caption{
Layer dependence of the magnitude of the self-consistent pair
potentials $|\Delta_{0,n}|$ in a 20-layer slab for the chiral
$p$-wave and $d_{x^2-y^2}$-wave channels. The chiral $p$-wave order
parameter is strongly enhanced at the outermost layers and decays
within only a few layers toward the interior, whereas the $d$-wave
order parameter remains negligibly small on this scale for all layers.
}
\label{fig:delta}
\end{figure}

Figure~\ref{fig:delta} displays the magnitude $|\Delta_{0,n}|$ as a
function of the layer index $n$ for both pairing channels in a
20-layer slab. In the chiral $p$-wave case, the order parameter
exhibits a pronounced enhancement at the outermost layers and decays
rapidly towards the interior of the slab, becoming essentially zero in
the central region. The spatial extent of the superconducting region is
limited to only two to three layers from each surface, indicating that
the pairing is strongly localized near the boundaries where the
drumhead states reside. In contrast, the $d$-wave order parameter
remains negligibly small on the scale of Fig.~\ref{fig:delta} for all
layers; its maximal amplitude is more than an order of magnitude
smaller than that of the chiral $p$-wave state. This pronounced
difference shows that, for the same interaction strength $V$, the
chiral $p$-wave channel is much more effective in developing a finite
order parameter than the $d$-wave channel. Because the $d$-wave gap
amplitude is so strongly suppressed, we do not analyze its quasiparticle
spectrum and LDOS in further detail.

%%%%%%%%%%%%%%%%%%%%%%%%%%%%%%%%%%%

\begin{figure}[t]
\centering
\begin{minipage}[t]{0.3\columnwidth}
    \centering
    \includegraphics[width=0.9\columnwidth]{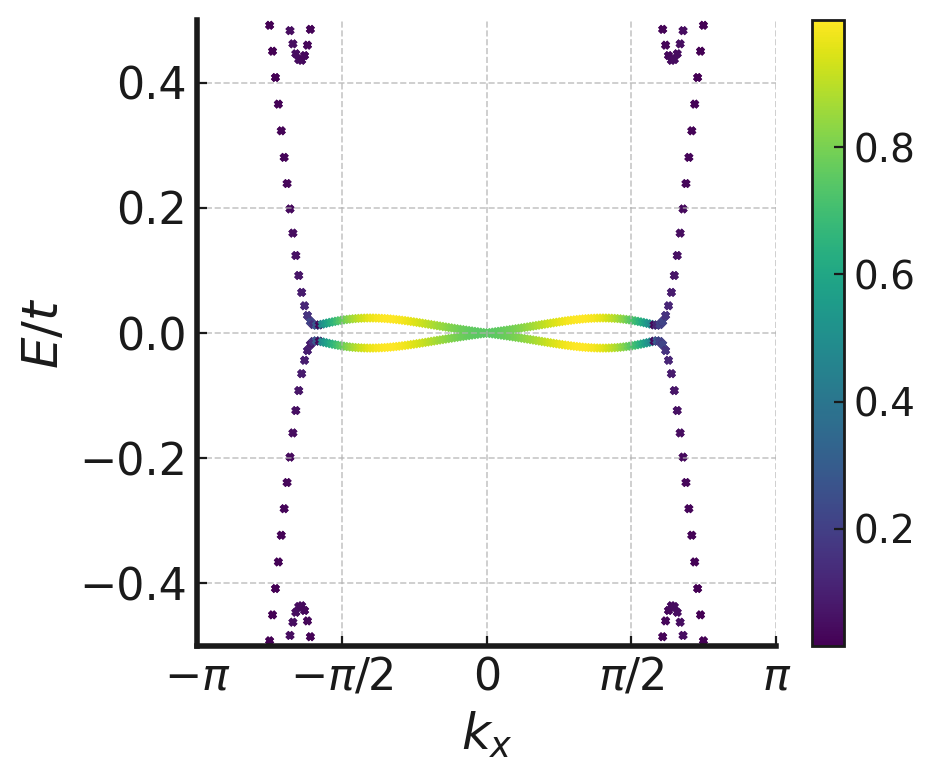}
    \subcaption{}
    \label{fig:band_SC}
\end{minipage}
\begin{minipage}[t]{0.3\columnwidth}
    \centering
    \includegraphics[width=0.9\columnwidth]{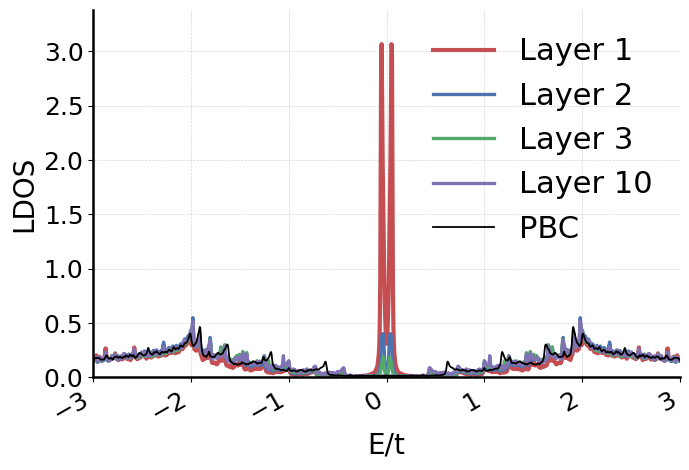}
    \subcaption{}
    \label{fig:LDOS_SC_3}
\end{minipage}
\begin{minipage}[t]{0.3\columnwidth}
    \centering
    \includegraphics[width=0.9\columnwidth]{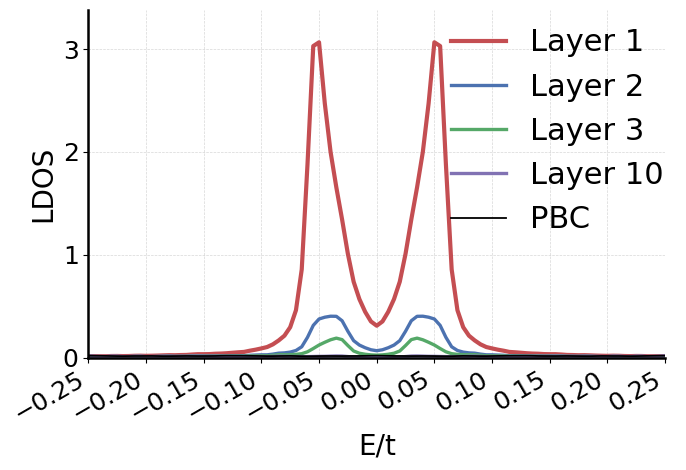}
    \subcaption{}
    \label{fig:LDOS_SC_0.25}
\end{minipage} 
\caption{
(a) Quasiparticle dispersion $E(k_x)$ along $k_y=0$ in the
self-consistent chiral $p$-wave state for a 20-layer slab.
(b) Local density of states on the surface layer in the same
state.
(c) Magnified view of the surface LDOS near $E=0$, showing the
splitting of the zero-energy peak originating from the drumhead
surface band and the resulting superconducting gap.
} \label{fig:SC}
\end{figure}

We next examine the spectral properties of the self-consistent chiral
$p$-wave state. Figure~\ref{fig:SC}\subref{fig:band_SC} shows the quasiparticle band
dispersion as a function of $k_x$ along the cut $k_y=0$ for the
20-layer slab. In the normal state [cf.~Fig.~\ref{fig:normal}\subref{fig:band_OBC}], the
drumhead surface band forms an almost flat dispersion at $E=0$ in the
interval $-2\pi/3 \lesssim k_x \lesssim 2\pi/3$. In the chiral
$p$-wave state, this flat band is pushed away from zero energy and a
superconducting gap opens over most of this momentum range: the
spectrum remains gapless only at $k_x=0$, whereas for
$k_x \neq 0$ inside the projected nodal ring the quasiparticle bands are
separated from the Fermi level by a finite gap.

The corresponding LDOS on the surface layer in the chiral $p$-wave
state is displayed in Fig.~\ref{fig:SC}\subref{fig:LDOS_SC_3}. Compared with the
normal-state surface LDOS in Fig.~\ref{fig:normal}\subref{fig:LDOS_normal}, where the
drumhead band gives rise to a sharp zero-energy peak, the spectral
weight around $E=0$ is strongly suppressed and redistributed into two
peaks at finite energies. This behavior is more clearly seen in
Fig.~\ref{fig:SC}\subref{fig:LDOS_SC_0.25}, which shows a magnified view of the
surface LDOS near $E=0$. The zero-energy peak originating from the
drumhead surface band is split into two peaks located symmetrically
about the Fermi energy, and the energy separation between these peaks
provides a direct measure of the superconducting gap induced on the
surface states.

These results demonstrate that the chiral $p$-wave order parameter that
develops near the surfaces efficiently gaps out the drumhead band,
leading to a well-defined superconducting gap in the surface spectra,
while leaving gapless excitations only at isolated momenta such as
$k_x=0$. Combined with the strong surface enhancement of
$|\Delta_{0,n}|$ shown in Fig.~\ref{fig:delta}, this confirms that,
within the present minimal model, superconductivity originating from the
drumhead surface states is strongly biased toward a chiral $p$-wave
pairing symmetry. The resulting superconducting state is confined to a
few layers near the surfaces, whereas the interior of the slab remains
essentially normal.

When the chemical potential is shifted away from zero,
the magnitude of the order parameter \(|\Delta_{0,n}|\)
gradually decreases. For \(\mu = 0.2\), it already falls
below the convergence threshold on all but the outermost
surface layer, indicating that the surface-localized
superconductivity is strongly suppressed once the Fermi
level is moved away from the bottom of the drumhead band.

%%%%%%%%%%%%%%%%%%%%%%%%%%%%%%%%%%%%%%%%%%%%%%%%%

%%%%%%%%%%%%%%%%%%%%%%%%%%%%%%%%%%%
\section{Summary}
%%%%%%%%%%%%%%%%%%%%%%%%%%%%%%%%%%%
In this work we have investigated surface superconductivity in a
nodal-line semimetal by using a minimal tight-binding model and a
layer-resolved Bogoliubov--de Gennes approach. In the normal state, we
confirmed that the model realizes a nodal loop in the bulk band
structure and the associated drumhead surface band in a slab geometry
with open boundaries in the $z$ direction. The central layers of a
20-layer slab faithfully reproduce the bulk-like density of states,
whereas the surface layer exhibits a pronounced zero-energy peak due to
the drumhead surface states.

On top of this nodal-line band structure we introduced superconductivity in two pairing channels, a chiral $p$-wave
state and a $d_{x^2-y^2}$-wave state, and determined the
layer-dependent gap amplitudes self-consistently. The chiral $p$-wave
order parameter develops predominantly on the outermost layers and
decays within only a few layers towards the interior, while remaining
essentially zero in the central region. In contrast, the $d$-wave order
parameter is strongly suppressed on all layers and its maximal amplitude
is more than an order of magnitude smaller than that of the chiral
$p$-wave state. The quasiparticle dispersion and surface local density
of states in the chiral $p$-wave state show that the drumhead surface
band is efficiently gapped out, and the zero-energy peak in the
normal-state surface spectrum is split into two coherence peaks that
directly reflect the induced superconducting gap. These results
demonstrate that superconductivity originating from the drumhead surface
states is strongly biased toward a surface-localized chiral $p$-wave
pairing symmetry, while the bulk interior of the slab remains essentially
normal. Our findings provide a simple microscopic scenario for
surface-dominated chiral superconductivity in nodal-line materials and
may offer qualitative guidance for interpreting surface-sensitive
experiments on Pd-doped CaAgP.

%%%%%%%%%%%%%%%%%%%%%%%%%%%%%%%%%%%
\section{Acknowledgements}
%%%%%%%%%%%%%%%%%%%%%%%%%%%%%%%%%%%
The numerical computations were carried out at the Cyber-science Center, Tohoku University.

%%%%%%%%%%%%%%%%%%%%%%%%%%%%%%%%%%%%%%%%%%%%
\bibliographystyle{iopart-num}
\bibliography{refs}

@article{CaAgP_nodal,
  author  = {Yamakage, Ai and Yamakawa, Youichi and Tanaka, Yukio and Okamoto, Yoshihiko},
  title   = {{Line-Node Dirac Semimetal and Topological Insulating Phase in Noncentrosymmetric Pnictides CaAgX (X = P, As)}},
  journal = {Journal of the Physical Society of Japan},
  volume  = {85},
  number  = {1},
  pages   = {013708},
  year    = {2016},
  doi     = {10.7566/JPSJ.85.013708},
  eprint  = {1510.00202},
  archivePrefix = {arXiv}
}

@article{CaAgP_transport,
  author  = {Okamoto, Yoshihiko and Saigusa, Kazushige and Wada, Taichi and Yamakawa, Youichi and Yamakage, Ai and Sasagawa, Takao and Katayama, Naoyuki and Takatsu, Hiroshi and Kageyama, Hiroshi and Takenaka, Koshi},
  title   = {{High-mobility carriers induced by chemical doping in the candidate nodal-line semimetal CaAgP}},
  journal = {Physical Review B},
  volume  = {102},
  pages   = {115101},
  year    = {2020},
  doi     = {10.1103/PhysRevB.102.115101},
  eprint  = {2008.06188},
  archivePrefix = {arXiv}
}

@inproceedings{CaAgP_pointcontact,
  author  = {Matsubara, Naoki and Nagasaka, Shota and Yano, Rikizo and Saigusa, Kazushige and Shinoda, Yusaku and Okamoto, Yoshihiko and Takenaka, Koshi and Kashiwaya, Satoshi},
  title   = {{Two-Band Model and Point Contact Spectroscopy of Nodal-Line Semimetal CaAg\textsubscript{0.9}Pd\textsubscript{0.1}P}},
  booktitle = {JPS Conference Proceedings},
  volume  = {38},
  pages   = {011028},
  year    = {2023},
  doi     = {10.7566/JPSCP.38.011028}
}

@article{CaAgP_TRSB1,
  author  = {Matsubara, Naoki and Yano, Rikizo and Saigusa, Kazushige and Takenaka, Koshi and Okamoto, Yoshihiko and Tanaka, Yukio and Kashiwaya, Satoshi},
  title   = {{Broken time-reversal symmetry detected by tunneling spectroscopy of superconducting Pd-doped CaAgP}},
  journal = {arXiv e-prints},
  year    = {2024},
  eprint  = {2412.08335},
  archivePrefix = {arXiv},
  primaryClass = {cond-mat.supr-con}
}

@article{CaAgP_TRSB2,
  author  = {Matsubara, Naoki and Yano, Rikizo and Saigusa, Kazushige and Takenaka, Koshi and Okamoto, Yoshihiko and Tanaka, Yukio and Kashiwaya, Satoshi},
  title   = {Broken time-reversal symmetry detected by point-contact spectroscopy of superconducting Pd-doped CaAgP},
  journal = {Physical Review B},
  volume  = {112},
  pages   = {014504},
  year    = {2025},
  doi     = {10.1103/xwds-q3w8}
}

@article{Shapourian2018,
  author  = {Shapourian, Hassan and Wang, Yuxuan and Ryu, Shinsei},
  title   = {Topological crystalline superconductivity and second-order topological superconductivity in nodal-loop materials},
  journal = {Physical Review B},
  volume  = {97},
  number  = {9},
  pages   = {094508},
  year    = {2018},
  doi     = {10.1103/PhysRevB.97.094508},
  eprint  = {1711.02122},
  archivePrefix = {arXiv},
  primaryClass = {cond-mat.supr-con}
}

@article{Trama2025,
  author  = {Trama, Mattia and K{\"o}nye, Viktor and Fulga, Ion Cosma and van den Brink, Jeroen},
  title   = {Self-consistent surface superconductivity in time-reversal symmetric Weyl semimetals},
  journal = {Physical Review B},
  volume  = {112},
  number  = {6},
  pages   = {064514},
  year    = {2025},
  doi     = {10.1103/PhysRevB.112.064514}
}
%%%%%%%%%%%%%%%%%%%%%%%%%%%%%%%%%%%%%%%%%%%%
\end{document}